\documentclass[12pt,preprint]{aastex}
\usepackage{graphicx}

\newcommand\beq{\begin{equation}}
\newcommand\eeq{\end{equation}}
\newcommand\lsim{\mathrel{\rlap{\lower4pt\hbox{\hskip1pt$\sim$}}
        \raise1pt\hbox{$<$}}}
\newcommand\gsim{\mathrel{\rlap{\lower4pt\hbox{\hskip1pt$\sim$}}
        \raise1pt\hbox{$>$}}}

\begin{document}
\title{Magnetic Scaling Laws for the Atmospheres of Hot Giant
  Exoplanets}

\author{Kristen Menou\altaffilmark{1}} \altaffiltext{1}{Department of
Astronomy, Columbia University, 550 West 120th Street, New York, NY
10027}

\begin{abstract}

We present scaling laws for advection, radiation, magnetic drag and
ohmic dissipation in the atmospheres of hot giant exoplanets. In the
limit of weak thermal ionization, ohmic dissipation increases with the
planetary equilibrium temperature ($T_{\rm eq} \gsim 1000$~K) faster
than the insolation power does, eventually reaching values $\gsim 1\%$
of the insolation power, which may be sufficient to inflate the radii
of hot Jupiters.  At higher $T_{\rm eq}$ values still, magnetic drag
rapidly brakes the atmospheric winds, which reduces the associated
ohmic dissipation power. For example, for a planetary field strength
$B=10$~G, the fiducial scaling laws indicate that ohmic dissipation
exceeds $1\%$ of the insolation power over the equilibrium temperature
range $T_{\rm eq} \sim 1300$--$2000$~K, with a peak contribution at
$T_{\rm eq} \sim 1600$~K. Evidence for magnetically dragged winds at
the planetary thermal photosphere could emerge in the form of reduced
longitudinal offsets for the dayside infrared hotspot. This suggests
the possibility of an anticorrelation between the amount of hotspot
offset and the degree of radius inflation, linking the atmospheric and
interior properties of hot giant exoplanets in an observationally
testable way.  While providing a useful framework to explore the
magnetic scenario, the scaling laws also reveal strong parameter
dependencies, in particular with respect to the unknown planetary
magnetic field strength.
\end{abstract}

\section{Introduction}

Hot giant exoplanets, including hot Jupiters, are among the best
characterized exoplanets. Secondary eclipses, transmission
spectroscopy and orbital phase curves in particular have provided
constraints on the atmospheric properties of several transiting
members of this class (see Charbonneau 2009; Winn 2010 for
reviews). Increasingly detailed radiative and circulation models for
the atmospheres of these exoplanets have also been developed to help
interpret the rich emerging phenomenology (see Burrows \& Orton 2010;
Showman et al. 2010 for reviews). Finally, evolutionary models have
been systematically used to infer the bulk interior properties of hot
giant exoplanets with precisely measured transit radii (see Baraffe et
al. 2010 for a review).

It was recently proposed that magnetic effects could have important
consequences for the atmospheric and bulk interior properties of hot
giant exoplanets. Magnetic drag and ohmic dissipation, which result
from kinematic induction by thermally-driven winds in the
weakly-ionized atmospheres of hot giant exoplanets,\footnote{We are
  excluding from our analysis giant exoplanets which are hot because
  of their young age. While magnetic effects may also be important for
  them, our focus is on older planets with atmospheric temperatures
  and winds determined by strong external irradiation, as exemplified
  by hot Jupiters.}  will indeed brake atmospheric winds (Perna et
al. 2010a; Rauscher \& Menou 2011) and deposit extra heat in the deep
atmosphere or the interior adiabat (Batygin \& Stevenson 2010; Perna
et al. 2010b), thus affecting some of the observable properties of
these exoplanets.

Here, we establish simple scaling laws for the strength of magnetic
drag and ohmic dissipation in the atmospheres of hot giant exoplanets,
as a function of their radiative equilibrium temperature. These
scaling laws allow us to study the nature of the transition into the
magnetized regime, as the radiative equilibrium temperature is raised
above $\sim 1000$~K, and to explore how the limit of strong coupling
with the planetary magnetic is eventually reached, at higher
temperatures. The remainder of this paper is organized as follows. In
\S2, we recall the basic mechanism of magnetic induction in the
atmosphere of a hot giant exoplanet. In \S3, we establish scaling laws
for the strength of magnetic drag and ohmic dissipation as a function
of the planetary radiative equilibrium temperature. The main results
emerging from these scaling laws are described in \S4. In \S5, we
discuss various parameter dependencies, the strength of induced
magnetic fields, connections to previous work on this topic and a few
plausible observational signatures of magnetic effects in the
population of hot giant exoplanets. We conclude in \S6.

\section{Basic Mechanism}

Day-side insolation on a hot Jupiter generates fast atmospheric winds
(see, e.g., reviews by Showman et al. 2008; 2010). As the
weakly-ionized gas flows across the planetary magnetic field, which is
presumably anchored to the planet's bulk rotation via deep-seated
electric currents in the convective interior, an additional magnetic
field is induced in the atmosphere. In steady-state, this magnetic
induction is balanced by resistive diffusion and associated ohmic
dissipation in the resistive atmosphere.

Let us consider a spherical coordinate system ($r,~\theta,~\phi$) that
is rotating with the planet. Let us further assume that the planetary
magnetic field takes the form of an aligned dipole (of surface
strength, $B_{\rm dip}$) and that the dominant atmospheric flow is
zonal (azimuthal) in nature: $V_{\phi} \gg V_r,\, V_{\theta}$. To
leading order, the induced current $J$ satisfies the steady-state
resistive induction equation
\begin{equation}
\frac{\partial {\bf B}}{\partial t} = \nabla \times ({\bf V_{\phi}} \times {\bf B_{\rm dip}}) - \nabla \times \left( \frac{4 \pi \eta}{c} { \bf J} \right)=0,
\label{eq:induc}
\end{equation}
where $\eta$ is the local resistivity of the weakly-ionized
atmospheric gas. Note the implicit assumption that the deep-seated
electric currents generating the dipolar planetary field are located
well below the atmospheric region of interest, so that they can be
omitted from the above induction equation (Liu et al. 2008; Batygin \&
Stevenson 2010; Perna et al. 2010a,b).

For a purely zonal flow and a strictly aligned dipole, a purely
toroidal field ($B_\phi$) is induced, which is sustained by purely
poloidal currents: ${\bf J }= (c /4 \pi) {\bf \nabla \times B_\phi}$ (see,
e.g., Liu et al. 2008).  Magnetic induction does not operate at the
rotational equator for an aligned dipole because vertical speeds are
negligibly small and meridional wind speeds do not contribute. The
aligned dipole assumption is likely to remain qualitatively correct
even for moderate levels of misalignment.

Focusing on the dominant, zonal component of the atmospheric flow, a
representative horizontal momentum balance for the steady flow in the
azimuthal (${\bf e_\phi}$) direction may be written as
\begin{equation}
\left( {\bf V_{\phi} \nabla  V_{\phi}} \right) .~{\bf e_\phi} = \left( -\frac{1}{\rho} {\bf \nabla} P - \frac{1}{\rho c}~ {\bf J} \times {\bf B_{\rm dip}}\right) .~{\bf e_\phi},
\label{eq:moment}
\end{equation}
where the zonal acceleration (LHS) resulting from the day-night
pressure gradient (first term on the RHS) is reduced by the bulk
Lorentz force due to magnetic drag (second term on the RHS; e.g., Zhu
et al. 2005).

Our main goal with this work is to study how this momentum balance
changes with the atmospheric resistivity, which depends itself
strongly on the planet's radiative equilibrium temperature via thermal
ionization balance. Considering the above induction and momentum
equations together, one anticipates the following behavior. At low
enough atmospheric temperatures (high resistivity $\eta$), magnetic
drag ($\propto J$) will not significantly reduce the zonal wind speed,
$V_{\phi}$ (in Eq.~[\ref{eq:moment}]). At fixed $V_{\phi}$ and $B_{\rm
  dip}$ values, the induced current $J$ will thus increase in inverse 
proportion to the resistivity $\eta$ (in Eq.~[\ref{eq:induc}]). At
large enough temperatures (low resistivity $\eta$), however, the winds
will be strongly dragged and it is no longer obvious how wind speeds
and induced currents should scale with resistivity in
Eq.~(\ref{eq:induc}). It is thus of great interest to establish the
range of atmospheric temperatures over which such a transition occurs
and to clarify what happens in the limit of high conductivity (high
temperatures).

\section{Magnetic Scaling Laws}

Our goal in establishing magnetic scaling laws for hot giant exoplanet
atmospheres is not to obtain an accurate description of magnetic
effects for this class of planets, as this will require detailed
multi-dimensional models. Rather, our focus is on the leading-order
scaling with the radiative equilibrium temperature, which sets the
atmospheric resistivity. As we establish these scaling laws, we will
need to invoke important simplifying assumptions.

\subsection{General Properties}

We adopt fiducial parameters for the scaling laws that represent a
typical hot Jupiter, with a hydrogen-dominated, solar composition
atmosphere, so that a perfect gas constant ${\cal R}=4.59 \times
10^7$~erg/g/K and a specific heat at constant pressure $C_p=1.43
\times 10^8$~erg/g/K are appropriate. The planetary radius is
$R_p=10^{10}$~cm, with a surface gravity $g=890$~cm~s$^{-2}$. We
explore surface magnetic field values from $B_{\rm dip}=3$~G to
$B_{\rm dip}=30$~G. Planetary radiative equilibrium temperatures
ranging from $T_{\rm eq}=1000$~K to $2200$~K are considered, which
correspond to semi-major axes $a \simeq 0.09$--$0.02$~AU around a
Sun-like star.\footnote{The radiative equilibrium temperature is
  defined by $T_{\rm eq}=T_* \left( \frac{R_*}{2D} \right)^{1/2}$ for
  zero Bond albedo, where $T_*$ and $R_*$ are the stellar effective
  temperature and radius, and $D$ is the planet-star orbital
  separation (e.g., Guillot 2010).}

\subsection{Model Atmospheres}

For the purpose of the present study, we focus our discussion on the
most actively forced, ``weather layer'' region of the atmosphere,
defined here as the region located between the thermal (infrared)
photosphere above and the insolation (visible) photosphere below,
which corresponds to a level above which $\sim 50\%$ of the insolation
flux has been absorbed (see details below). For simplicity, the
vertical thermal structure of the atmosphere is approximated with the
one-dimensional radiative solution discussed by Guillot (2010; see
also Hansen 2008) for a zero incidence angle and a dilution factor
$f=0.5$ representing a dayside average. As shown by Guillot (2010; see
his Fig.~4), this solution provides reasonable temperature-pressure
profiles for hot Jupiter atmospheres over a range of orbital radii,
even though it clearly is a simplified approach and it neglects
important three-dimensional aspects of the atmospheric structure. In
what follows, we refer to this pressure-dependent temperature profile
for the dayside weather layer as $T_{\rm day}$. Note that, by
construction, $T_{\rm day} \gsim T_{\rm eq}$ everywhere in this
weather layer model.

Following Guillot (2010), a single thermal absorption coefficient
$\kappa_{\rm th}= 10^{-2}$~cm$^2$~g$^{-1}$ and a single visible
absorption coefficient $\kappa_{\rm v}= 4 \times
10^{-3}$~cm$^2$~g$^{-1}$ are adopted. For these values, the thermal
and visible photospheres (defined by $\tau=2/3$) are located,
respectively, at $P_{\rm th} \simeq 60$~mbar and $P_{\rm v} \simeq
150$~mbar (above which $\sim 50\%$ of the stellar flux has been
absorbed). The number of pressure scale heights over which the weather
layer extends vertically in our models is thus modest, $\Delta \ln P =
\ln P_{\rm v} - \ln P_{\rm th} \simeq 0.9$. An internal heat flux
corresponding to $T_{int}=150$~K is adopted for the thermal structure
calculation, but the exact value has little effect on the atmospheric
properties of the weather layer of interest here.

Guillot (2010) has also shown that these values for the grey
absorption coefficients are adequate to reproduce the global radiative
equilibrium balance of the prototypical hot Jupiter HD209458b, with
$T_{\rm eq} \simeq 1420$~K. The fixed values of the absorption
coefficients adopted in our scaling laws, despite the large range of
$T_{\rm eq}=1000$--$2200$~K considered, is arguably one of the
strongest simplifying assumptions we make in our models. We comment on
the effects of variations in opacities and other model parameters in
\S\ref{sec:depend}.

\subsection{Ionization Balance}

Thermal ionization, rather than photoionization, is expected to be
dominant over the range of atmospheric pressures ($P \gsim 50$~mbar)
of interest for the scaling laws (see, e.g., Perna et al. 2010a;
Batygin \& Stevenson 2010). Rather than using the same approximate
solution to the Saha equation as in Perna et al. (2010a,b), which is
only justified at low enough temperatures ($T \lsim 1800$~K), we use
here a more general but still simplified solution to the Saha equation
(e.g., Sato 1991). In the weakly-ionized regime, when only single
ionization has to be considered, the ionization fraction $x_e ~(\ll
1)$ satisfies the equations

\begin{eqnarray}
x_e & = & \frac{n_e}{n_n}= \sum_j \frac{n_j}{n} x_j,\\
\frac{x_j^2}{1- x_j^2} & \simeq &\frac{1}{n_j k T} \left( \frac{2 \pi m_e}{h^2} \right)^{3/2} (kT)^{5/2} \exp(-I_j/kT),\nonumber
\label{eq:xe}
\end{eqnarray}

where $n_e$ and $n_n$ are the number densities of electrons and
neutrals, respectively (in cm$^{-3}$), $n_j$ is the number density of
element $j$, $I_j$ is the corresponding first ionization potential,
$T$ is the temperature, $n$ is the total number density of the gas and
the other symbols have their usual meaning. Solar composition for the
first 28 elements of the periodic table (H to Ni) is assumed for the
ionization calculations. In the limit $x_j \ll 1$, which ceases to be
valid for some elements at $T \gsim 2000$~K for the densities of
interest here, the above expression essentially reduces to that used
by Laughlin et al. (2011). Based on this solution for the ionization
balance, the local electric resistivity of the atmospheric gas is
evaluated everywhere in the weather layer as $\eta= 230 \sqrt{T}
/x_e~{\rm cm^2~s^{-1}}$ (see Perna et al. 2010a for details).

The ionization solutions obtained with Eq.~(\ref{eq:xe}) are still
approximate, however, since all the degeneracy factors in the Saha
equation have been arbitrarily set to unity. This crude simplification
is justified by the fact that the specific atmospheric composition of
any given hot Jupiter is a priori unknown and could easily deviate
from solar in a non-trivial way. In addition, from the point of view
of the scaling laws, inaccuracies in the degeneracy factors are
dwarfed by the exponential dependence with temperature appearing in
Eq.~(\ref{eq:xe}), given the wide range of temperatures
considered. Indeed, we verified that factor several changes in the
degeneracy factors entering the Saha equation have little practical
impact on our scaling law results.

\subsection{Forcing, Momentum Balance and Dragged Wind Speeds} \label{sec:forc}

Showman et al. (2010) describe how to obtain a simple,
order-of-magnitude scaling for the zonal wind speed in a hot
Jupiter atmosphere, in the absence of magnetic drag, based on steady
non-linear balance between the zonal acceleration term and the
pressure gradient term in the horizontal momentum equation. The same
dimensional analysis of Eq.~(\ref{eq:moment}), taking the limit
$B_{dip}=0$, leads to the same scaling
\begin{equation}
V_{\phi} = \sqrt{ {\cal R}~ \Delta T_{\rm horiz} ~\Delta \ln P},
\end{equation}

where $\Delta T_{\rm horiz}$ is the typical day-night temperature
differential along the equator, ${\cal R}$ is the perfect gas constant
and $\Delta \ln P$ measures the number of vertical pressure scale
heights in the weather layer over which this horizontal temperature
differential applies. 

In the presence of magnetic drag, a similar dimensional analysis of
Eq.~(\ref{eq:moment}) leads to the balance
\begin{equation}
\frac{V_{\phi}^2}{R_p} = \frac{{\cal R}~ \Delta T_{\rm horiz} ~\Delta \ln P}{R_p} - \frac{V_{\phi} B^2}{4 \pi \rho \eta},
\label{eq:est_v}
\end{equation}
where the bulk Lorentz force has been evaluated via dimensional
analysis of the steady-state resistive induction equation
(Eq.~[\ref{eq:induc}]) for the latitudinal current: $J_\theta\sim
c{\rm V}_\phi B/(4\pi\eta)$ (see, e.g., Perna et al. 2010a).  We
obtain an order-of-magnitude estimate of the zonal wind speed in the
presence of magnetic drag by solving for the positive root of this
quadratic equation for $V_{\phi}$.

An additional complication with the above estimate is that the
day-night temperature differential itself, $\Delta T_{\rm horiz}$,
depends on the zonal wind speed $V_\phi$ via the efficiency of heat
advection from the dayside to the nightside. This results in a
non-linear coupling between advection and radiation processes in the
atmospheric flow. While Showman \& Guillot (2002) proposed to capture
this dependence with an exponential relation between $\Delta T_{\rm
  horiz}$ and the ratio of the atmospheric advective and radiative
timescales, we adopt a power-law dependence here for simplicity,
keeping the index as a free parameter. The horizontal temperature
differential between the day and the night sides of a tidally-locked,
hot giant exoplanet is thus estimated as
\begin{equation}
\Delta T_{\rm horiz} = \min \left[\frac{T_{\rm day}}{2}, \frac{T_{\rm day}}{2} \left( \frac{\tau_{\rm adv}}{\tau_{\rm rad}} \right)^n \right],
\label{eq:est_dt}
\end{equation}
so that when $\tau_{\rm adv}/\tau_{\rm rad} >1$, the full differential
$T_{\rm day}/2$ is applied, while if $\tau_{\rm adv}/\tau_{\rm rad}
<1$, advection reduces the effective temperature differential between
day and night in proportion to the values of these two timescales.  A
default value $n=1$ is adopted for the power law index but other
values are explored below.

The advective timescale is evaluated as

\beq
\tau_{\rm adv} = \frac{R_p}{V_{\phi}},
\eeq

while the radiative timescale is evaluated as

\beq 
\tau_{\rm rad} = \frac{C_p P}{g \sigma T_{\rm day}^3}, \label{eq:trad}
\eeq

where $g$ is the gravitational acceleration, $P$ and $T_{\rm day}$ are
the pressure and dayside temperature at the location of interest in
the weather layer and $\sigma$ is the Stefan-Boltzmann constant (e.g.,
Goody \& Yung 1989).

Equations~(\ref{eq:est_v}) and~(\ref{eq:est_dt}) for our two unknowns,
$V_\phi$ and $\Delta T_{\rm horiz}$, are solved iteratively. As it
turns out, for the majority of the parameter space of interest here,
atmospheres are strongly radiative so that the precise form of the
scaling with $\tau_{\rm adv}/\tau_{\rm rad} $ or $n$ in
Eq.~(\ref{eq:est_dt}) has little consequences for our main
conclusions.

The ohmic power dissipated per unit volume in relation to the magnetic
drag is evaluated as

\beq
\label{eq:Qohm}
Q_{\rm ohm}= \frac{4 \pi \eta J^2}{c^2}=\frac{V_{\phi}^2 B^2}{4 \pi \eta}.
\eeq

Integrating vertically, and then horizontally over the volume of the
weather layer, this yields an estimate of the total ohmic power
dissipated as a result of induction in the resistive atmosphere,

\beq
P_{\rm ohm} = 4 \pi R_p^2  \int_{\Delta \ln P} Q_{\rm ohm} \times  H_p ~d \ln P,
\eeq

where the local pressure scale height is evaluated as $H_p= {\cal R}
T_{\rm day} / g$. It is worth emphasizing that this estimate of the
ohmic power is not expected to be accurate at better than an order of
magnitude level, given that it overlooks important issues related to
the geometry and the spatial distribution of electrical currents
within and outside the weather layer (see, e.g., Liu et al. 2008;
Batygin \& Stevenson 2010; Perna et al. 2010b).

\section{Results}

\begin{figure*}
\centering \includegraphics[scale=0.65]{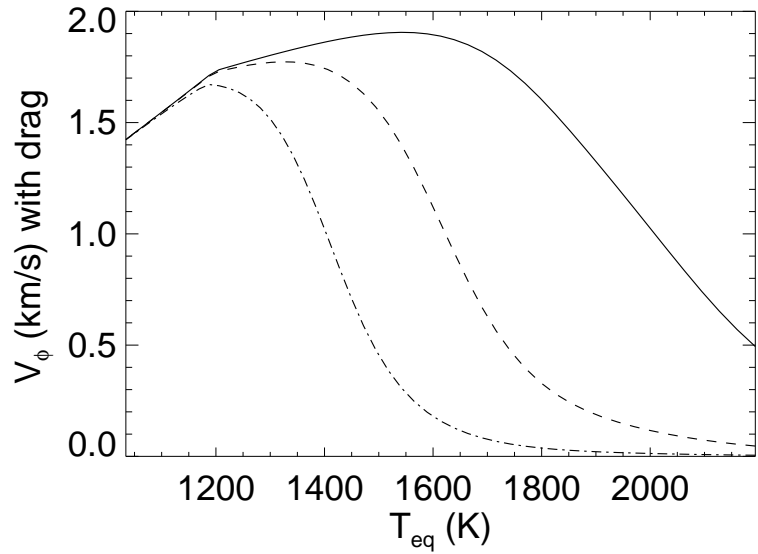}
\caption{Estimate of the zonal wind speed, $V_{\phi}$ (in km/s), at
  the thermal photosphere as a function of the planetary radiative
  equilibrium temperature, $T_{\rm eq}$ (in K). Solid, dashed and
  dash-dotted lines show results for surface magnetic field strengths
  $B_{\rm dip} =3$, $10$ and $30$~G, respectively. The exponential
  decline of $V_{\phi}$ at high $T_{\rm eq}$ is caused by strong
  magnetic drag.}
\label{fig:one}
\end{figure*}

Figure~\ref{fig:one} shows our estimate for the zonal wind speed,
$V_{\phi}$ at the thermal photosphere ($P_{\rm th} \simeq 60$~mbar) as
a function of the planetary radiative equilibrium temperature, $T_{\rm
  eq}$. Results for an assumed magnetic field strength $B_{\rm dip}
=3$, $10$ and $30$~G are shown as solid, dashed and dash-dotted lines,
respectively.

These curves cover three different regimes for momentum balance in the
thermally-forced atmospheres of hot giant exoplanets. At low $T_{\rm
  eq}$ values, advection reduces the day-night temperature
differential $\Delta T_{\rm horiz}$ according to
Eq.~(\ref{eq:est_dt}), which leads to the relatively steep slope of
$V_\phi$ with $T_{\rm eq}$. As shown best by the solid line, at higher
$T_{\rm eq}$ values, this slope becomes shallower as the atmosphere
becomes radiatively dominated (when $\tau_{\rm adv}/\tau_{\rm rad} >
1$), with a negligible role for advection (as defined by
Eq.~[\ref{eq:est_dt}]). Finally, the three curves show that magnetic
drag eventually becomes efficient at braking the winds and that it
does so at lower radiative equilibrium temperatures for stronger
magnetic field strengths.  Results comparable to those shown in
Fig.~\ref{fig:one} are obtained for deeper levels in the weather layer
than the thermal photosphere.

\begin{figure*}
\centering \includegraphics[scale=0.65]{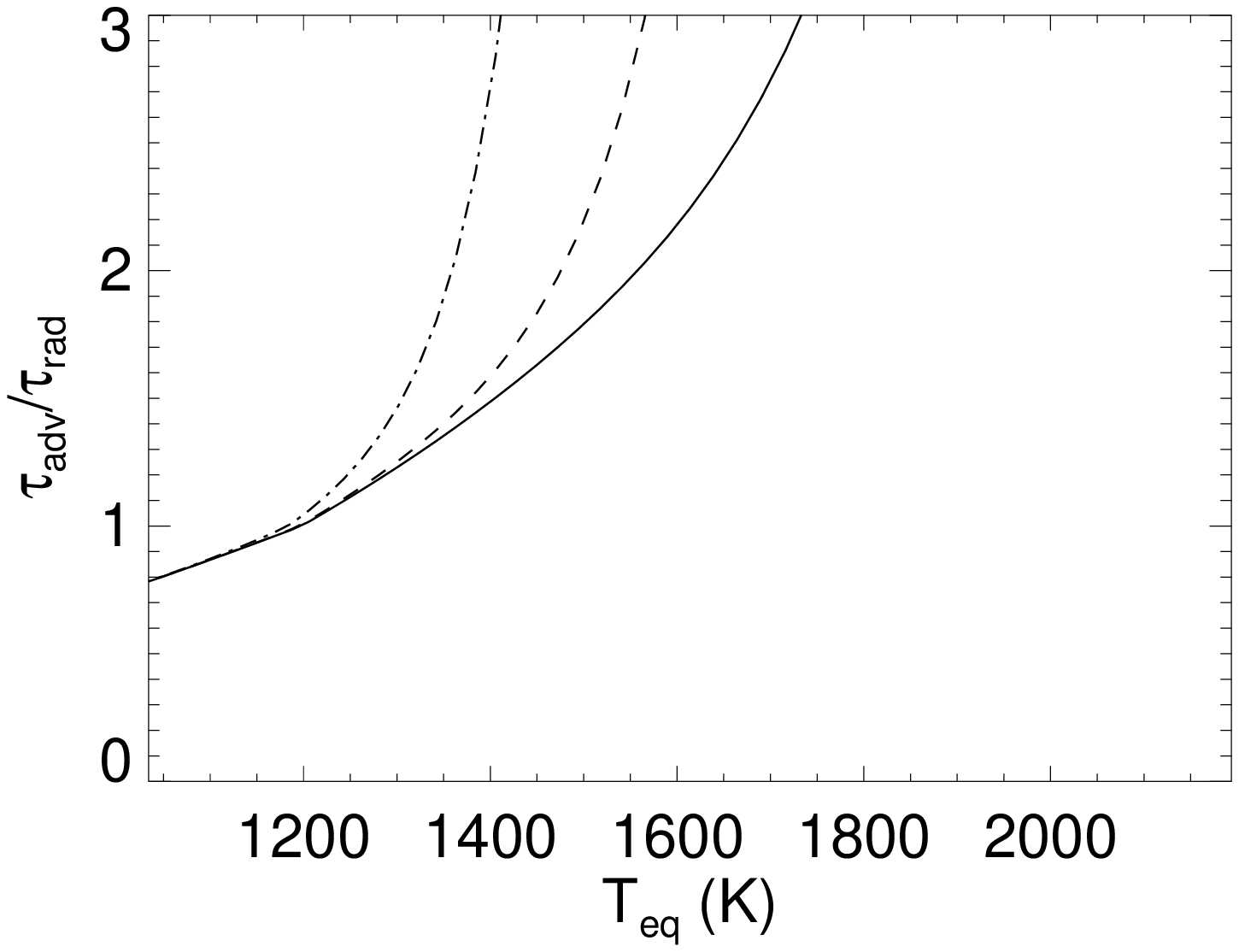}
\caption{Ratio of advective to radiative timescales, $\tau_{\rm
    adv}/\tau_{\rm rad}$, at the thermal photosphere as a function of
  the planetary radiative equilibrium temperature, $T_{\rm eq}$ (in
  K). Solid, dashed and dash-dotted lines show results for surface
  magnetic field strengths $B_{\rm dip} =3$, $10$ and $30$~G,
  respectively. The radiative equilibrium temperature above which
  advection becomes negligible ($\tau_{\rm adv}/\tau_{\rm rad} >1$) is
  comparable to the temperature at which the strong magnetic drag
  regime begins for $B_{\rm dip} \sim 30$~G.  }
\label{fig:onebis}
\end{figure*}

\begin{figure*}
\centering \includegraphics[scale=0.65]{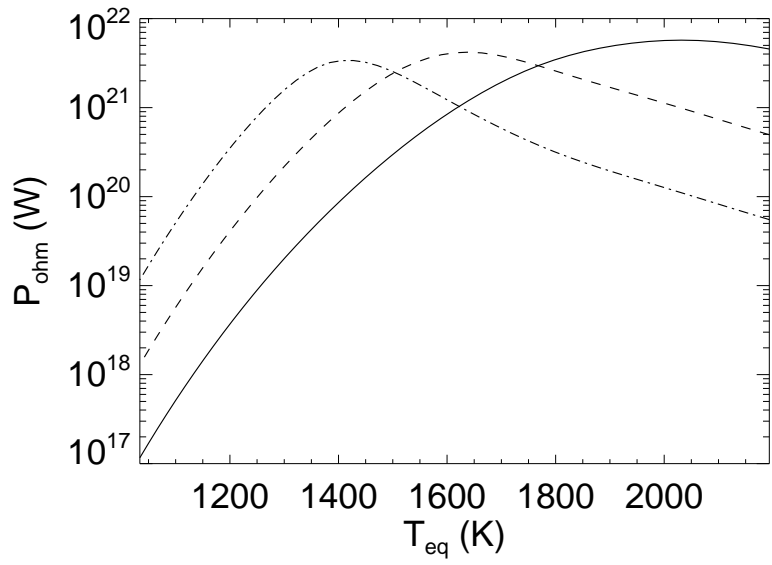}
\caption{Ohmic power (in Watts) dissipated as a result of induction in
  the weather layer as a function of the planetary radiative
  equilibrium temperature, $T_{\rm eq}$ (in K). Solid, dashed and
  dash-dotted lines show results for surface magnetic field strengths
  $B_{\rm dip} =3$, $10$ and $30$~G, respectively. Ohmic power rises
  exponentially with $T_{\rm eq}$ until the regime of strong magnetic
  drag is reached, at which point the power starts declining
  exponentially.}
\label{fig:two1}
\end{figure*}

\begin{figure*}
\centering \includegraphics[scale=0.65]{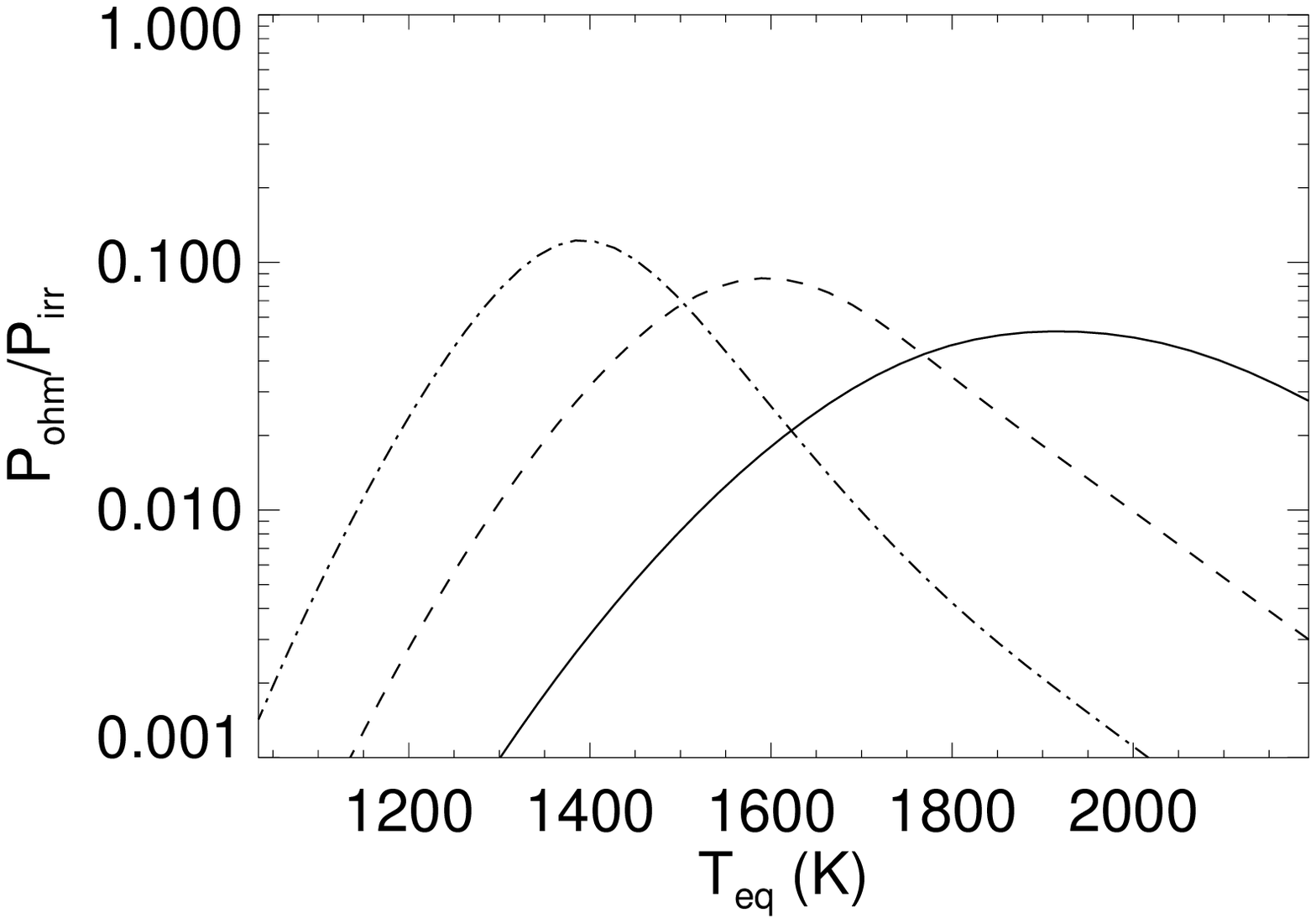}
\caption{Ratio of ohmic to irradiation power ($P_{\rm ohm}/P_{\rm
    irr}$) as a function of the planetary radiative equilibrium
  temperature, $T_{\rm eq}$ (in K). Solid, dashed and dash-dotted
  curves show results for surface magnetic field strengths $B_{\rm
    dip} =3$, $10$ and $30$~G, respectively. Assuming that a ratio
  $P_{\rm ohm}/P_{\rm irr} > 1$\% is required for sizable planetary
  radius inflation, this figure illustrates how radius inflation
  should only exist over a restricted range of equilibrium
  temperatures, according to our magnetic scaling laws.}
\label{fig:two2}
\end{figure*}

The exponential decline in $V_{\phi}$ at high $T_{\rm eq}$ values,
which is caused by strong magnetic drag, can be understood as
follows. As resistivity drops, the dominant balance in the momentum
equation (Eq.~[\ref{eq:est_v}]) eventually becomes one where the
pressure gradient acceleration term is balanced by the deceleration
term from the bulk Lorentz force. In that limit, $V_\phi$ becomes
small and simple dimensional analysis in that limit shows that it
scales in proportion to the resistivity, $V_\phi \propto \eta$, which
itself is an exponentially decreasing function of the atmospheric
temperatures in the weather layer, and thus of $T_{\rm eq}$.

Figure~\ref{fig:onebis} illustrates the nature of the change from an
advective to a radiative behavior for hot giant exoplanet atmospheres
as the radiative equilibrium temperature, $T_{\rm eq}$, is raised
above $\sim 1200$~K. Below $1200$~K, $\tau_{\rm adv}/\tau_{\rm rad}
<1$ and advection reduces the day-night temperature differential
according to Eq.~(\ref{eq:est_dt}). Above $1200$~K, however, the
atmospheric thermal structure (at the thermal photosphere) becomes
essentially radiative, with little role, if any, for advection. As
shown by the three curves in Fig.~\ref{fig:onebis}, this transition is
effectively accelerated by magnetic drag, when it significantly
reduces the wind speeds, the more so for strong magnetic field
strengths.

Figure~\ref{fig:two1} shows the ohmic power $P_{\rm ohm}$ dissipated
as a result of induction in the weather layer as a function of the
planetary radiative equilibrium temperature, $T_{\rm eq}$.  The
exponential rise and subsequent exponential decline of $P_{\rm ohm}$
with $T_{\rm eq}$ can be understood as follows. In the limit of weak
magnetic drag, $V_\phi$ is independent of the resistivity and it
varies only weakly with $T_{\rm eq}$. Equation~(\ref{eq:Qohm}) then
implies that $P_{\rm ohm} \propto Q_{\rm ohm} \propto 1/\eta$. On the
other hand, in the limit of strong magnetic drag, since $V_\phi
\propto \eta$, the ohmic power declines exponentially with $T_{\rm
  eq}$ according to $P_{\rm ohm} \propto Q_{\rm ohm} \propto \eta$. As
shown by Fig.~\ref{fig:two1}, the radiative equilibrium temperature at
which the ohmic power peaks depends sensitively on the value of the
surface magnetic field strength.

Figure~\ref{fig:two2} shows a different representation of the ohmic
dissipation results, where the ohmic power $P_{\rm ohm}$ has been
scaled to the total irradiation power $P_{\rm irr}= 4 \pi R_P^2 \sigma
T_{\rm eq}^4$ received by a hot giant exoplanet with radiative
equilibrium temperature, $T_{\rm eq}$. The ratio $P_{\rm ohm}/P_{\rm
  irr}$ is useful because it has been suggested that a value $\gsim
1$\% is needed for sizable radius inflation of a hot
Jupiter,\footnote{We note that any energetic threshold for radius
  inflation, such as the ratio $P_{\rm ohm}/P_{\rm irr}$, will
  generally be a function of the planet's mass and age (e.g., Miller
  et al. 2009), so that the $1$\% value adopted here should be taken
  as merely indicative.}  by deposition either in the bulk convective
interior (Batygin \& Stevenson 2010) or in the deep atmosphere (Perna
et al. 2010b). Our simple magnetic scaling laws lack the descriptive
power to identify where in the planet the ohmic power is dissipated,
but they clarify the range of radiative equilibrium temperatures over
which such power is in principle available for radius inflation. Like
the peak ohmic value, the range of radiative equilibrium temperatures
over which $P_{\rm ohm}/P_{\rm irr} > 1$\% (or any other fixed ratio)
is a sensitive function of the surface magnetic field strength.

\section{Discussion}

\subsection{Parameter Dependencies}
\label{sec:depend}

An examination of the various assumptions made in deriving our
magnetic scaling laws makes it clear that they can only capture the
most basic, order-of-magnitude behavior for hot giant exoplanet
atmospheres across the wide range of radiative equilibrium
temperatures considered. It is important for the validity of the
scaling laws that the central physical ingredient of the theory,
atmospheric resistivity, varies exponentially with atmospheric
temperatures, via thermal ionization balance. Still, there is some
freedom in renormalizing the scaling laws by adopting different
parameter values than the fiducial ones we have adopted. It is thus
important to understand the sensitivity of the scaling law results to
variations in the different model parameters.

We find that modest changes ($\sim 20\%$) in the assumed values of the
planetary radius, $R_p$, the gravitational acceleration, $g$, the
perfect gas constant, ${\cal R}$, or the specific heat at constant
pressure, $C_p$, have only minor quantitative effects on the scaling
law results. Raising the assumed gas metallicity to ten times solar,
which increases the free electron density accordingly, reduces the
temperature at which the strong drag regime is entered (from, e.g.,
$\sim 1600$~K to $\sim 1475$~K, for $B_{\rm dip}=10$~G), but the
scaling law results are otherwise qualitatively similar to the solar
metallicity case.  Similarly, changing the power law index used to
evaluate the day-night temperature differential in
Eq.~(\ref{eq:est_dt}) from $n=1$ to $n=3$ or $n=5$ further steepens
the slope of $V_\phi$ with $T_{\rm eq}$ at low $T_{\rm eq}$ values
(leftmost region in Fig.~\ref{fig:one}), but the scaling law results
remain otherwise qualitatively similar.

The scaling law results are sensitive to variations in model
parameters which directly affect the atmospheric temperatures in the
weather layer, however, as expected from the exponential dependence of
the resistivity with temperature. For example, increasing the thermal
opacity coefficient, $k_{\rm th}$, by $50$\% nearly doubles the ohmic
power and shifts the ohmic peak from $\sim 1600$~K to $\sim 1500$~K,
for $B_{\rm dip}=10$~G. Reducing $k_{\rm th}$ by a factor of two leads
to about three times less ohmic dissipation, a peak ohmic power at
$\sim 1650$~K for $B_{\rm dip}=10$~G and magnetic Reynolds numbers
reduced by a factor of a few. Increasing the visible opacity
coefficient, $k_{\rm v}$, by $50$\% reduces the ohmic power by a
factor $\sim 5$, shifts the ohmic peak to slightly lower temperatures
and reduces magnetic Reynolds numbers by a factor of a few. Reducing
the value of $k_{\rm v}$ by a factor of two increases the ohmic power
by a factor $\sim 5$, shifts the ohmic peak to slightly larger
temperatures and increases magnetic Reynolds numbers by a factor of a
few.

Similarly, an increase in the assumed dayside dilution factor, $f$,
entering Guillot's radiative solution raises the ohmic power somewhat
and shifts its peak to lower temperature (from, e.g, $\sim 1600$~K for
$f=0.5$ to $\sim 1475$~K for $f=0.7$, for $B_{\rm dip}=10$~G). A
reduction in the magnitude of the maximum day-night temperature
differential in Eq.~(\ref{eq:est_dt}) from $0.5~ T_{\rm day}$ to $0.3~
T_{\rm day}$ reduces the ohmic power by about a factor two and the
peak $V_\phi$ values by $\sim 25$\%. Conversely, an increase of this
maximum day-night temperature differential from $0.5~ T_{\rm day}$ to
$0.7~ T_{\rm day}$ doubles the ohmic power and increases the peak
$V_\phi$ value by $\sim 20$\%.

Another factor which may significantly impact atmospheric temperatures
in the weather layer, and thus the magnetic scaling laws, is the
possibility of a temperature inversion in the upper atmosphere of
strongly-irradiated giant exoplanets (e.g., Fortney et al. 2008). Hot
giant exoplanets with such inversions would typically have reduced
temperatures in their weather layer, from extra absorption of stellar
light at altitude. Even though we did not explore this scenario in any
detail, and instead adopted a simple set of Guillot (2010) radiative
solutions with $\kappa_{\rm th} >\kappa_{\rm v}$ for our thermal
profiles, one can anticipate a systematic shift of the magnetic
scaling law results at fixed $T_{\rm eq}$ for those planets with
inversions, in proportion to the temperature deficit in their weather
layer. A detailed investigation of the effects of temperature
inversions on magnetic scaling laws would constitute a valuable
extension of our work.

Besides the strong dependence on the planetary magnetic field
strength, which is illustrated explicitly in Figs.~1-4, our parameter
space exploration reveals a strong dependence of the scaling law
results on parameters affecting the atmospheric temperatures in the
modeled weather layer. This suggests that improved treatments of the
atmospheric thermal structure, e.g. based on three-dimensional
modeling, would greatly benefit efforts to better understand the
magnetic behavior of hot giant exoplanet atmospheres as a function of
their radiative equilibrium temperature.

\subsection{Magnetic Reynolds Number and Induced Fields}

As the radiative equilibrium temperature of a hot giant exoplanet is
raised, the degree of coupling between the atmospheric flow and the
planetary magnetic field also increases.  The nature of this coupling
in the strong drag limit identified by our scaling laws, at the high
end of the range of equilibrium temperatures considered, is an
important issue.

\begin{figure*}
\centering \includegraphics[scale=0.65]{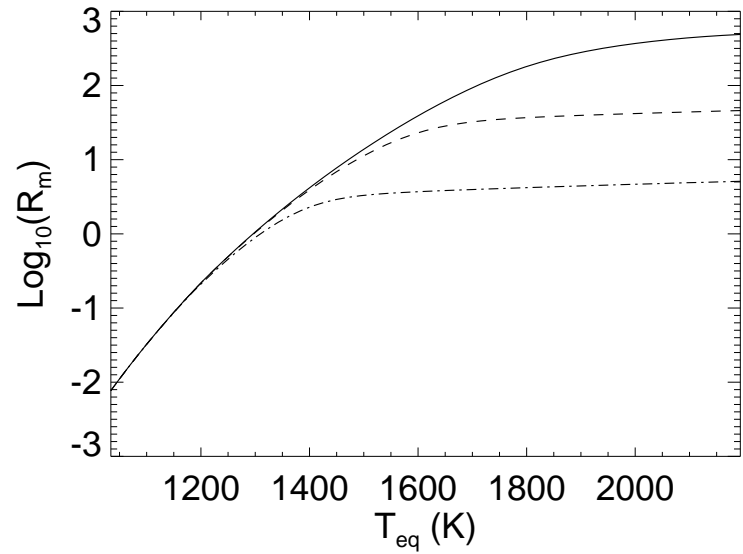}
\caption{Logarithm of the magnetic Reynolds number, $R_{\rm m}$, of
  the atmospheric flow at the thermal photosphere, as a function of
  the planetary radiative equilibrium temperature, $T_{\rm eq}$ (in
  K). Solid, dashed and dash-dotted lines show results for a surface
  magnetic field strength $B_{\rm dip} =3$, $10$ and $30$~G,
  respectively. The exponential drop in $V_{\phi}$ caused by strong
  magnetic drag is responsible for the near-asymptotic values of
  $R_{\rm m}$ reached at high $T_{\rm eq}$.  }
\label{fig:four}
\end{figure*}

Figure~\ref{fig:four} shows an estimate of the magnetic Reynolds
number of the atmospheric flow, $R_m$, as a function of the planetary
radiative equilibrium temperature, $T_{\rm eq}$, using the same
notation as in previous figures. This magnetic Reynolds number is
evaluated as
\begin{equation}
R_m= \frac{V_\phi H_p}{ \eta},
\end{equation}
where $V_\phi$ is the zonal wind velocity at the thermal photosphere
(Fig.~\ref{fig:one}), $H_p$ is the pressure scale height and $\eta$ is
the resistivity. Fig.~\ref{fig:four} reveals that the magnetic
Reynolds number is less than unity and independent of the magnetic
field strength at low enough radiative equilibrium
temperatures. Beyond $T_{\rm eq} \sim 1300$~K, however, $R_m$ exceeds
unity and reaches different, near-asymptotic values for different
strengths of the planetary magnetic field. The near-asymptotic
behavior of $R_m$ at high temperatures can easily be understood as
resulting from the previously established scaling $V_\phi \propto
\eta$ in the strong drag limit, with different normalizations for the
different magnetic field strengths.

Traditionally, the regime $R_m \gsim 1$ would indicate a significant
degree of coupling between the flow and the magnetic field, and the
possibility of sustained dynamo action in the atmosphere. It is
unclear, however, whether the moderately large values of $R_m$ found
in Fig.~\ref{fig:one} are sufficient to trigger a self-sustained
dynamo, and what the nature of such a dynamo might be, given the
unusual setup of a very shallow atmospheric flow threaded by the
planetary magnetic field.

Irrespective of the issue of sustained dynamo action, the significant
gas-field coupling arising from values of $R_m \gsim 1$ raises the
possibility of large induced field strengths even in the simple
kinematic framework adopted in this study. In particular, a strong
enough induced toroidal magnetic field ($B_{\phi}$) could in principle
result in secondary induction of a strong poloidal field component by
meridional atmospheric motions. This could challenge one of our basic
assumption, which is that a dominant contribution to the poloidal
field component is provided by the planetary dipole field ($B_{\rm
  dip}$ in Eq.~[\ref{eq:induc}]).

Some insight on this issue may be gained by simple dimensional
analysis. Using the same scaling as in \S\ref{sec:forc} for the
latitudinal current, $J_\theta\sim c{\rm V}_\phi B_{\rm dip}
/(4\pi\eta)$, and assuming that the relevant gradient scale is $\sim
H_p$ in the relation between the induced current and the induced
field, ${ J_\theta = (c /4 \pi) \nabla \times B_\phi}$, one deduces a
simple estimate of the strength of the induced toroidal field in terms
of the background dipole field: $B_\phi \sim R_m B_{\rm dip}$. In the
strong drag limit, this suggests values for the induced toroidal field
which can be in excess of the planetary dipole field.

If large enough field values were induced into a poloidal component by
meridional motions, Eq.~(\ref{eq:induc}), which is a starting point
for our scaling laws, could cease to be valid: the value of the
poloidal field may no longer be set by the fixed value of the
planetary dipole field but rather by multi-dimensional induction in
the atmospheric flow itself. It is unclear, however, whether such
large induction of poloidal field is expected in the weather layer of
hot giant exoplanets. Indeed, simple dimensional analysis analogous to
the one performed above suggests an induced poloidal field strength
$B_{\rm pol} \sim R_m^{*} B_\phi$, where $ R_m^{*} = V_{\theta}
H_p/\eta$ is the magnetic Reynolds number associated with the
meridional (rather than zonal) flow. Therefore, to the extent that the
meridional flow velocities ($V_{\theta}$) are much weaker than the
zonal ones, in the presence of separate drags acting on each
component, $R_m^{*}$ may still be small enough for the induced
poloidal field to remain negligible compared to the background dipole
field.

Our scaling law results in the strong drag limit assume that the
planetary dipole field remains dominant relative to any induced
poloidal field. It is difficult to evaluate if or when this assumption
might break down without considerably more detailed explorations of
the physics of magnetized atmospheric flows than performed here, such
as multi-dimensional models with distinct drag treatments for the
zonal and meridional flows, or perhaps even full MHD treatments to
explore dynamo issues. Until such studies become available, it will
thus be important to remain cautious when using simple scaling
arguments like ours to interpret the phenomenology of hot giant
exoplanet atmospheres in the strong drag limit.

\subsection{Connection to Previous Work}

Our scaling law results are broadly consistent with previously
published results for the zonal wind speeds in the weather layer of
hot giant exoplanets. In particular, the values shown in
Fig.~\ref{fig:one} for zonal wind speeds are qualitatively consistent
with the zonally-averaged results of atmospheric circulation models
for the hot giant exoplanet HD209458b ($T_{\rm eq} \sim 1400$~K) with
weak ($B_{\rm dip}=3$~G), moderate ($B_{\rm dip}=10$~G) and strong
($B_{\rm dip}=30$~G) imposed drag, as reported in Perna et
al. (2010a). Extrapolating the trend at low $T_{\rm eq}$ values in
Fig.~\ref{fig:one} also points to zonal wind speeds in broad agreement
with those reported by Lewis et al. (2010) for the atmosphere of the
hot Neptune GJ436b ($T_{\rm eq} \sim 650$~K) and it supports the
notion that the atmospheric flow on this cooler exoplanet can
effectively be treated as free of magnetic drag.

The nature of the transition from the unmagnetized to the magnetized
atmosphere regime for hot giant exoplanets is a particularly
interesting issue to examine in light of our scaling law results. In
terms of magnetic drag, Fig.~\ref{fig:one} indicates that the
magnetized regime starts around $T_{\rm eq} \sim 1200$--$1600$~K,
depending on the strength of the planetary magnetic field. Defining
the value of $T_{\rm eq}$ above which ohmic dissipation begins to be
important is more difficult because it depends on a threshold value
for the ratio of ohmic to irradiation power above which significant
radius inflation is expected. This ratio is not well defined because
it depends on details of the ohmic dissipation process but a threshold
value of $P_{\rm ohm}/P_{\rm irr} \sim 1$\% is probably a reasonable
guess (Guillot \& Showman 2002; Batygin \& Stevenson 2010; Perna et
al. 2010b). As shown in Fig.~\ref{fig:two2}, this $1$\% threshold is
reached at $T_{\rm eq} \sim 1100$--$1500$~K according to our scaling
laws, with a strong dependence on the planetary magnetic field
strength. When evaluating the transition in the magnetized regime, one
should remember the great degree of simplification used in deriving
our scaling laws, which is exemplified by some of the strong parameter
dependencies discussed in \S\ref{sec:depend}. The nature of the
spatial distribution of ohmic dissipation will also likely influence
the transition temperature for the onset of radius inflation (Batygin
\& Stevenson 2010; Perna et al. 2010b).

In their study of the heavy element content of giant exoplanets,
Miller \& Fortney (2011) suggest a temperature of $ \sim 1000$~K as
the threshold below which radius inflation appears to be absent from
their sample of exoplanets. This temperature threshold is broadly
consistent with, but somewhat lower than, the threshold values
inferred on the basis of our scaling law results, shown in
Fig.~\ref{fig:two2}. It may be possible to bring these threshold
values into closer agreement by lowering the critical value of $P_{\rm
  ohm}/P_{\rm irr}$ required for sizable radius inflation (below our
$1\%$ assumption) and/or by postulating that planetary magnetic fields
in the exoplanet sample considered by Miller \& Fortney (2011) are
generally strong ($B_{\rm dip} \gsim 30$~G).

In an analysis sharing some similarities with the present one,
Laughlin et al. (2011) have studied a correlation between a measure of
radius inflation for hot giant exoplanets, the so-called radius
anomaly, and their effective temperature,\footnote{For
  strongly-irradiated planets with negligible internal heat fluxes,
  effective and radiative equilibrium temperatures are equivalent.} in
an observationally constrained sample. The emergence of a positive
radius anomaly at $T_{\rm eff} \gsim 1000$~K, as shown in their
Fig.~2, is broadly consistent with our scaling law results. It is
unclear whether or not our work supports the power law dependence of
the radius anomaly with $T_{\rm eff}$ advocated by Laughlin et
al. (2011). This relation depends on the intricate physics that
connects ohmic dissipation with the effective radius inflation and it
simply is not addressed by our scaling laws. It is conceivable,
however, that the large spread in the amount of radius anomaly found
at any $T_{\rm eff}$ value (see their Fig.~2) could be accounted for
by variations in the strengths of the planetary magnetic field in the
sample, since we find $B_{\rm dip}$ to be a rather sensitive parameter
of the theory.  We note that there also is an interesting indication
of a systematic drop in radius anomaly at $T_{\rm eff} \gsim 1900$~K
in the Fig.~2 of Laughlin et al. (2011). It is possible that this
feature is associated with the drop in ohmic dissipation, and thus in
the amount of radius inflation, that is expected in the strong drag
regime according to our scaling law results (see our
Fig.~\ref{fig:two2}).

Batygin et al. (2011) have studied in detail some of the consequences
of ohmic dissipation for the thermal evolution of hot giant
exoplanets. In their analysis, these authors assume that the global
efficiency, $\epsilon$, of ohmic heating relative to that from stellar
irradiation is a model constant, fixed at values of $1$, $3$ or
$5$\%. Our scaling law results challenge this simplifying assumption
since we find that the ratio $P_{\rm ohm}/P_{\rm irr}$ is a strong
function of both the planetary effective temperature, $T_{\rm eff}$,
and the planetary magnetic field strength (see our
Fig.~\ref{fig:two2}). The emergence of sizable radius inflation at
$T_{\rm eff} \gsim 1400$~K in the evolutionary models of Batygin et
al. (2011; see their Figs 6--8) is certainly consistent with the
typical threshold values suggested by our scaling laws, but the
existence of a well-defined transition temperature would require a
rather uniform set of planetary magnetic strengths according to our
analysis. Given the difference in methodology, we note that it may be
possible to recast some of the results of Batygin et al. (2011) in the
language of our scaling laws, by considering that their assumed ohmic
efficiency values of $\epsilon=1$, $3$ and $5$\% correspond to
different effective planetary magnetic field strengths.

Batygin et al. (2011) argued that magnetic drag has a very limited
effect on the magnitude of zonal winds in hot giant exoplanet
atmospheres. This is inconsistent with our scaling law results, which
do suggest the existence of a strong drag regime, as shown in our
Fig.~\ref{fig:one}, and it also conflicts with the strongly dragged
simulation results described in Perna et al. (2010b) and Rauscher \&
Menou (2011). Our work suggests that the emergence of a strong drag
regime is essential to the saturation and the eventual decline of
ohmic dissipation at high planetary effective temperatures. It is
worth noting that a systematic reduction in the efficiency of ohmic
dissipation $\epsilon$ in the models of Batygin et al. (2011) when
$T_{\rm eff} \gsim 1600$--$1900$~K would likely bring their model
radii in closer agreement with the data (see their Figs.~6-8). Such a
drop of $\epsilon$ with $T_{\rm eff}$ could also mitigate the
emergence of unstable, overflowing evolutionary tracks identified by
Batygin et al. (2011) for models with high enough values of $T_{\rm
  eff}$ ($\gsim 1400$--$1800$~K).

\subsection{Observational Signatures}

Magnetic drag and ohmic dissipation are separate manifestations of the
same magnetic induction process operating in the atmosphere of a hot
giant exoplanet. It is of great interest to explore the type of
observational signatures expected from magnetic drag and ohmic
dissipation since these observables should ultimately be related to
each other. As discussed earlier, the most direct observable
manifestation of ohmic dissipation is planetary radius
inflation. Signatures of magnetic drag, on the other hand, would most
likely emerge in the form of reduced atmospheric wind speeds.

Over the past few years, it has become possible to constrain the
magnitude of wind advection in the atmosphere of a few hot giant
exoplanets, by measuring their infrared phase curves (e.g., Harrington
et al. 2006; Cowan et al. 2007; Knutson et al. 2007, 2009a,b;
Crossfield et al. 2010). The standard interpretation that has been
given to an infrared peak emission that is offset from the planet's
orbital phase is that heat is advected away from the dayside by
eastward equatorial winds at the planet's thermal photosphere (e.g.,
Showman et al. 2008, 2009, 2010; Rauscher \& Menou 2010; Dobbs-Dixon
et al. 2010; Burrows et al. 2010; Heng et al. 2011; Showman \& Polvani
2011).  Magnetic drag will act to reduce the speed of zonal winds and
thus the amount of heat advection away from the dayside, so that
atmospheres experiencing stronger magnetic drag may exhibit reduced
infrared phase offsets (see, e.g., Fig.~3 of Rauscher \& Menou 2011)

A simple measure of the degree of heat advection and infrared phase
offset is provided by the ratio $\tau_{\rm adv}/\tau_{\rm rad}$ shown
in Fig.~\ref{fig:onebis}. As $T_{\rm eq}$ increases, $\tau_{\rm rad}$
decreases steeply (see Eq.~[\ref{eq:trad}]) but ultimately it is the
exponential decline experienced by the zonal wind speed $V_\phi$ in
the strong drag limit that leads to the large values of $\tau_{\rm
  adv}/\tau_{\rm rad}$ at high temperatures in
Fig.~\ref{fig:onebis}. One would thus expect smaller infrared phase
offsets for giant exoplanets with higher radiative equilibrium
temperatures. It is difficult to evaluate the magnitude of such
offsets with the simple ratio $\tau_{\rm adv}/\tau_{\rm rad}$, as
opposed to detailed atmospheric circulation models, but we note that
HD189733b, with $T_{\rm eq} \simeq 1200$~K, still has a measurable
offset (Knutson et al. 2007, 2009a) for a value of $\tau_{\rm
  adv}/\tau_{\rm rad} \sim 1$ (according to Fig.~\ref{fig:onebis}; see
also discussion in Agol et al. 2010).

Interestingly, our scaling law results indicate that the radiative
equilibrium temperature at which ohmic dissipation peaks is comparable
to that at which magnetic drag reduces the zonal wind speed by about a
half (relative to the case without magnetic drag). A comparison
between Figures~\ref{fig:one} and~\ref{fig:two2} reveals that this
trend holds independently of the planetary magnetic field strength. It
suggests that, across a range of $T_{\rm eq}$ values, from the start
of the magnetized regime until the ohmic peak is reached (e.g., $\sim
1300$--$1600$~K for $B_{\rm dip}=10$~G), zonal wind speeds will
systematically drop as ohmic dissipation increases. Observationally,
this might translate into an anticorrelation between the magnitude of
infrared phase offsets and the degree of radius inflation of hot giant
exoplanets. At higher temperatures, beyond the ohmic peak, one would
expect both zonal wind speeds and ohmic dissipation to drop with
$T_{\rm eq}$. The large values of the ratio $\tau_{\rm adv}/\tau_{\rm
  rad} ~(\gg 1)$ expected at such high temperatures may prevent any
observable infrared phase offset, but the decline in ohmic dissipation
may result in a gradual disappearance of the radius inflation
phenomenon with $T_{\rm eq}$, as already mentioned above in connection
to Fig. 2 of Laughlin et al. (2011). Cowan \& Agol (2011) have also
noticed that hot Jupiters receiving the largest irradiation fluxes
(corresponding to $T_{\rm eq} \gsim 1900$~K) appear to be inefficient
at redistributing heat to their night sides, a trend which may be
consistent with the emergence of a strong drag regime.  Given the
idealized nature of our scaling laws, it will be interesting to put
these various arguments to the observational test and to interpret the
results with more detailed, possibly planet-specific atmospheric
models.

\section{Conclusion}

We developed scaling laws addressing the magnitude of magnetic drag
and ohmic dissipation in the atmospheres of hot giant exoplanets, as a
function of the planetary radiative equilibrium temperature. These
scaling laws shed light on the nature of the transition from an
unmagnetized to a magnetized regime of atmospheric behavior, at
$T_{\rm eq} \gsim 1000$~K, and on the emergence of a strong drag
regime at even higher temperatures. Assuming that poloidal field
induction remains weak relative to toroidal induction, an issue which
deserves further scrutiny, our scaling laws suggest that ohmic
dissipation eventually vanishes in the limit of very strong
drag. Clarifying the role of magnetic drag and ohmic dissipation for
hot giant exoplanets is important and we have suggested that their
interplay could manifest in the form of an anticorrelation between the
amount of infrared phase offset and the degree of radius inflation for
such planets, thus linking the atmospheric and interior properties of
hot giant exoplanets in an observationally testable way.

\acknowledgements

The author thanks Charles Gammie for providing the original ionization
balance routine used in this work. This work was supported in part by
NASA grants JPLCIT 1366188 and PATM NNX11AD65G.

\end{document}